\documentstyle[12pt,epsf,cite]{article}
\def\Journal#1#2#3#4{{#1} {\bf #2}, #3 (#4)}

\def\NPB{{\em Nucl. Phys.} {\bf B}}
\def\PLB{{\em Phys. Lett.} {\bf B}}
\def\PRL{\em Phys. Rev. Lett.}
\def\PRD{{\em Phys. Rev.} {\bf D}}

\def\PR{\em Phys. Rep.}

\def\APP{\em Astropart. Phys.}
\def\APJS{\em Astrophys. J. Suppl.}
\def\PL{{\em Phys. Lett.}}
\def\JHEP{{\em J. High Energy Phys.}}
\def\IJMP{{\em Int. J. Mod. Phys.}}
\def\JCAP{{\em J. Cosmo. Astropart. Phys.}}
\newcommand{\gev}{\,{\rm GeV}}
\newcommand{\tev}{\,{\rm TeV}}
\title{ \vspace*{-10mm}
\begin{tabular}{ll}
\hspace*{10cm} & {\normalsize hep-ph/0404100} \vspace{-3mm} \\
\hspace*{10cm} & {\normalsize April 2004}
\end{tabular} \vspace{1cm} \\
\bf 
Relic density and elastic scattering cross sections of the neutralino 
in the MSSM with CP-violating phases} 
\author{
Takeshi Nihei\footnote{E-mail: nihei@phys.cst.nihon-u.ac.jp} \,
 and 
Masaki Sasagawa\footnote{E-mail: masaki.sasagawa@nifty.com} \\
 \mbox{} \vspace{3mm}\\
\normalsize \em 
Department of Physics, College of Science and Technology, Nihon University, \\
\normalsize \em 
1-8-14, Kanda-Surugadai, Chiyoda-ku, Tokyo, 101-8308, Japan }
\date{ }
\begin{document}
\baselineskip 18pt
\renewcommand{\thefootnote}{\fnsymbol{footnote}}
\maketitle
\begin{abstract}
\normalsize
\baselineskip 18pt
We study the neutralino relic density and the neutralino-proton elastic scattering cross sections in the minimal supersymmetric standard model (MSSM) with CP-violating phases. We include all the final states to the neutralino pair annihilation cross section at the tree level, taking into account the mixing between the CP-even and CP-odd neutral Higgs fields. We show that variations of the relic density and the elastic scattering cross sections with the CP-violating phases are significant. In particular, interferences between the major contributions in the neutralino annihilation cross section can cause considerable enhancement of the relic density. 
\end{abstract}
\newpage
%
%
\section{Introduction}
%

Recent progress in cosmological observations has much impact on 
determination of various cosmological parameters. 
The analyses of the anisotropy in the cosmic microwave background radiation 
using the first year WMAP data provide the relic abundance of 
a cold dark matter (CDM) as \cite{WMAP1,WMAP2} 
\begin{eqnarray}
\Omega_{\rm CDM}h^2 & = & 0.1126^{+0.008}_{-0.009}, 
\label{eqn:WMAP}
\end{eqnarray}
where $\Omega_{\rm CDM}$ is the CDM energy density normalized by
the critical density and 
$h$ $\approx$ 0.7 is a parameter in the Hubble constant 
$H_0$ $=$ 100 $h$ km/sec/Mpc \cite{Hubble}. 
The current allowed region of the CDM relic density is expected to be 
narrowed by the analyses of increasing WMAP data and 
forthcoming data from the future project Planck \cite{PLANCK}.

One of the most promising candidates for the CDM is the lightest 
superparticle (LSP) in the minimal supersymmetric standard model (MSSM) 
with R-parity conservation \cite{MSSM}. 
The R-parity guarantees that the LSP is stable. 
In this model, the LSP is typically the lightest neutralino 
\cite{neutralino-dm} which is 
a linear combination of neutral gauginos and higgsinos
\begin{eqnarray}
\chi & = & N_{11} \tilde{B} + N_{12} \tilde{W}^3 
+ N_{13} \tilde{H}^0_1 + N_{14} \tilde{H}^0_2, 
\label{eqn:chi-1}
\end{eqnarray}
where $\tilde{B}$ is the $U(1)_Y$ gaugino (bino), and 
$\tilde{W}^3$ is the neutral $SU(2)_L$ gaugino (wino). 
$\tilde{H}^0_1$ and $\tilde{H}^0_2$ are the two neutral higgsinos
with opposite hypercharges. 
The coefficients $N_{1i}$ ($i$ $=$ $1,2,3,4$) are the elements of 
the 4$\times$4 unitary matrix $N$ which diagonalizes the neutralino 
mass matrix \cite{MSSM}.

The lightest neutralino has weak but finite couplings to the quarks. 
This implies that it may be possible to detect the neutralino CDM 
directly in a detector. 
There have been various ongoing and planned activities on the 
direct detection of Weakly Interacting Massive Particles 
(WIMPs) \cite{DAMA,CDMS,EDELWEISS,ZEPLIN}. 
Current experimental upper bounds on the elastic scattering cross section
of the WIMP off proton 
are around $10^{-6}$ pb and $1$ pb for spin-independent (SI) 
interactions and spin-dependent (SD) ones, respectively. 
Future experiments are expected to drastically improve the sensitivity.  
Thus we may observe a first signal of a superparticle 
in the WIMP searches.

There have been a lot of theoretical analyses on the direct detection 
\cite{Goodman-Witten,DD-old,Drees-Nojiri-dd,DD-intermediate,DD-recent,KNRR} 
and the relic density
\cite{SWO,Eli-Ros-Lal,Drees-Nojiri,rd-ww,Gondolo-Gelmini,Lopez-Nano-Yuan,Baer-Brhlik,NRR1,RRN,NRR2,oh2-recent}
of the neutralino CDM in the MSSM \cite{Jungman-elal}. 
However, because of a large number of free parameters in the MSSM, most 
analyses assume some pattern of the supersymmetric parameters to reduce
a number of free parameters, 
such as a common sfermion mass, a definite relation on the gaugino masses, 
ignoring generation mixing or complex phases in the supersymmetric parameters. 
Considering increasing sensitivities of experimental projects, 
it is now important to relax these conditions to examine the possible 
variations of the CDM observables under more general conditions.

In particular, the dependence of the CDM observables on the CP-violating 
complex phases needs to be carefully examined. 
It has been realized that non-zero phases of 
the trilinear scalar couplings associated with the third generation 
sfermions ($A_t$, $A_b$, $A_{\tau}$) 
or the Higgs mixing mass parameters ($\mu$) can induce 
the mixing between the CP-even and CP-odd neutral Higgs fields 
\cite{Pilaftsis,cp_even_odd_mixing_2}. 
In Ref.\cite{dd-with-CPV}, the direct detection cross sections of the
lightest neutralino in the presence of CP violation were studied 
ignoring this mixing. 
An analysis of the direct detection cross sections with supersymmetric 
CP violation taking into account the CP even--odd mixing in the
neutral Higgs sector has been done in Ref. \cite{CPJS}. 
It was shown that the CP-violating phases can reduce or 
enhance the neutralino-nucleus cross sections significantly.

On the other hand, 
a full analysis of the neutralino relic density $\Omega_{\chi} h^2$ 
including all the contributions to the annihilation cross section 
in the MSSM with CP violation is missing. 
The effect of CP violation on 
the neutralino annihilation cross section into the fermion pairs 
through the sfermion exchange was studied in Ref.\cite{FOS}. 
It was shown that a non-zero phase in the squark left-right mixing 
removes the p-wave suppression of the annihilation cross section for the 
fermion pair final states, and greatly enhances the cross section. 
In Ref. \cite{GF}, the effects of a CP-violaing phase in the trilinear 
scalar couplings on the neutralino pair annihilation cross section 
were examined, including all the tree level contributions and 
taking into account the mixing between the CP-even
and CP-odd neutral Higgs fields. 
It was found that, 
the annihilation cross section can be enhanced typically by 
factors of one to four; the enhancement can be huge 
near the resonances where the relic density 
increases by a large resonant contribution of 
the s-channel Higgs exchange. 
Recently, the dependence of $\Omega_{\chi} h^2$ on CP-violating 
phases through supersymmetric loop corrections to the bottom-quark 
mass was examined, taking into account the mixing between the CP-even
and CP-odd neutral Higgs fields and focussing on the sfermion exchange and 
the Higgs boson exchange contributions to the fermion pair production
\cite{GINS}. 
It was shown that, near the resonances, 
the relic density is sensitive to CP-violating phases 
mainly through the loop corrections to the bottom-quark mass.

In the present work, we re-investigate the effects of CP-violating 
complex phases of supersymmetric parameters on the neutralino relic 
density and the neutralino-proton elastic scattering cross sections. 
We include all the contributions to the neutralino annihilation cross 
section at the tree level, taking into account the mixing between 
the CP-even and CP-odd Higgs fields. 
We examine the effects of CP violation outside the resonance regions.

This paper is organized as follows. 
In section 2, we declare our assumption on the parameters in 
the MSSM with supersymmetric CP violation, and briefly describe the
structure of the relevant interactions. 
In section 3, we give the formulae on the direct detection cross
sections. 
In section 4, we briefly sketch our method to compute the relic density
of the lightest neutralino. 
In section 5, we present our numerical results. 
Finally we give conclusions in Section 6.

%
%
\section{The MSSM with supersymmetric CP violation}
%

In general, the MSSM has a lot of free parameters
\cite{MSSM}. 
In the prerent work,
we examine the effects of the following parameters 
\begin{eqnarray}
M_1, \ M_2, \ M_3, \ \mu, \ m_{\tilde{f}}, \ A_f,
\ \tan \beta, \ m_A, 
\label{eqn:mssm-parameters}
\end{eqnarray}
where $M_1$, $M_2$ and $M_3$ are the mass parameters for 
the bino, the wino and the gluino, respectively. 
$\mu$ is the Higgs mixing mass parameter.
$m_{\tilde{f}}$ is the supersymmetry breaking common mass parameter 
for the sfermions. 
$A_f$ is the common trilinear scalar coupling for the third generation, 
while the ones for the first two generations are neglected. 
$\tan \beta$ is the ratio 
of the vacuum expectation values of the two neutral Higgs fields. 
$m_A$ is a parameter which coincides with the CP-odd Higgs mass 
in the absence of CP violation \cite{Pilaftsis,cp_even_odd_mixing_2}. 
In eq.(\ref{eqn:mssm-parameters}), 
$M_i$ ($i$ $=$ $1,2,3$), 
$\mu$ and $A_f$ can have CP-violating phases in general.

For the gaugino masses, we consider two cases: 
grand unified theory-like (GUT-like) case and 
anomaly-mediated supersymmetry breaking-like (AMSB-like) case. 
For the GUT-like case, the absolute values of the gaugino masses
are related as 
\begin{eqnarray}
|M_1| & = & \frac{5}{3}\tan^2 \theta_W |M_2|, 
\ \ \ \ |M_3| \ = \ \frac{\alpha_s}{\alpha_2} |M_2|,
\label{eqn:gut-like}
\end{eqnarray}
where $\theta_W$ is the Weinberg angle, and $\alpha_2$ and $\alpha_s$ are
the gauge coupling constants for $SU(2)_L$ and $SU(3)_C$ gauge 
groups, respectively. 
In the GUT-like case, the neutralino LSP can be bino-like 
($|M_1|$ $<$ $|\mu|$) or higgsino-like ($|M_1|$ $>$ $|\mu|$). 
For the AMSB-like case, they are related as 
\cite{AMSB} 
\begin{eqnarray}
|M_1| & = & 2.8|M_2|,
\ \ \ \ |M_3| \ = \ -8.3|M_2|. 
\label{eqn:amsb-like}
\end{eqnarray}
In this case, the neutralino LSP can be wino-like 
($|M_2|$ $<$ $|\mu|$) or higgsino-like ($|M_2|$ $>$ $|\mu|$). 
Note that the SU(5) grand unification or the anomaly-mediated supersymmetry 
breaking scenario implies the similar relations  
not for the absolute values $|M_i|$ ($i$ $=$ $1, 2, 3$) 
but for $M_i$ themselves. 

We take a convention that $M_2$ is real. 
Also, $A_f$ and $M_3$ are assumed to be real. 
The CP-violating phases we examine in the present work
are the phases of $M_1$ and $\mu$
\begin{eqnarray}
M_1 & = & |M_1|\exp(i\theta_{M_1}), 
 \ \ \mu \ = \ |\mu|\exp(i\theta_{\mu}). 
\label{eqn:phase}
\end{eqnarray}

It has been realized that CP-violating phases in $A_f$ or $\mu$ 
can induce the mixing between the CP-even and CP-odd Higgs fields
through radiative corrections \cite{Pilaftsis,cp_even_odd_mixing_2}. 
The MSSM contains two CP-even neutral Higgses $\phi_1$ and $\phi_2$ 
with hypercharges $Y(\phi_1)$ $=$ $-Y(\phi_2)$ $=$ $-1/2$ and a CP-odd
neutral Higgs $a$ as physical fields. 
In general, non-zero phases in $A_f$ or $\mu$ induce $\phi_1$--$a$ and 
$\phi_2$--$a$ mixings at the one-loop level, so that 
the Higgs boson mass-squared matrix ${\cal M}_H^2$
becomes 3$\times$3. 
We calculate the Higgs mass-squared matrix using the one-loop effective 
potential, including the contributions of the third generation 
fermions and sfermions. 
The mass eigenstates $H^0_r$ ($r$ $=$ 1, 2, 3) are related with the 
CP eigenstates ($\phi_1$,$\phi_2$,$a$) by a 3$\times$3 rotation 
matrix $O_H$ as follows
\begin{eqnarray}
\left( 
\begin{array}{c}
H^0_1 \\ H^0_2 \\ H^0_3 
\end{array}
\right)
 & = & 
O_H 
\left( 
\begin{array}{c}
\phi_1 \\ \phi_2 \\ a
\end{array}
\right). 
\end{eqnarray}
The Higgs boson mass eigenvalues are obtained as $O_H {\cal M}_H^2 O_H^T$
$=$ diag($m_{H^0_1}^2$,$m_{H^0_2}^2$,$m_{H^0_3}^2$). 
Note that the parameter $m_A$ in eq.(\ref{eqn:mssm-parameters})
is not a mass eigenvalue in general. 
The CP even--odd mixing in the neutral Higgs sector implies 
$(O_H)_{i3}$ $\neq 0$ ($i=1,2$) so that interactions of the neutral Higgs
bosons are significantly modified\cite{Pilaftsis,cp_even_odd_mixing_2}. 

If the mass parameters $M_1$ and $\mu$ have non-zero phases, 
the elements of the neutralino mixing matrix $N$ have an imaginary part
in general. 
In this case, the interactions of the neutral Higgses with the lightest 
neutralino and the fermions ($f$) relevant for the present analysis
have the following structure 
\begin{eqnarray}
{\cal L}_{\chi\chi H^0} & = & 
\sum_{r=1}^3 \bar{\chi} 
  (C_S^{\chi\chi H^0_r} - C_P^{\chi\chi H^0_r} \gamma_5) \chi H^0_r, 
\label{eqn:int_h0_nn} \\
{\cal L}_{ff H^0} & = & 
\sum_{f=u,d,\cdots} \sum_{r=1}^3 \bar{f} 
(C_S^{ff H^0_r} - C_P^{ff H^0_r} \gamma_5) f H^0_r.
\label{eqn:int_h0_ff}
\end{eqnarray}
Without CP violation, either the scalar coupling $C_S$ or the 
pseudoscalar coupling $C_P$ is vanishing for every mass eigenstate 
$H^0_r$ \cite{MSSM}. 
In the presence of CP violation, however, 
both $C_S$ and $C_P$ are non-zero for every mass eigenstate. 
The neutralino--fermion--sfermion interactions have 
similar structure
\begin{eqnarray}
{\cal L}_{f\tilde{f}\chi} & = & 
\sum_{f=u,d,\cdots} \sum_{i=1,2} \bar{\chi}
(C_S^{\chi f\tilde{f}_i}- C_P^{\chi f\tilde{f}_i}\gamma_5 )f \tilde{f}_i
 \ + \ {\rm h.c.} \ .
\label{eqn:int_chi_f_sf}
\end{eqnarray}
We have neglected generation mixing. The two mass eigenstates
$\tilde{f}_i$ ($i$ $=$ $1,2$) for the sfermions are given by linear
combinations of the gauge eigenstates $\tilde{f}_L$ and $\tilde{f}_R$
with the same flavor. 
The expressions for the coupling constants in 
equations (\ref{eqn:int_h0_nn}), (\ref{eqn:int_h0_ff}) and
(\ref{eqn:int_chi_f_sf}) can be found in Ref \cite{CPJS,NRR1}.

%
\section{Direct detection of the neutralino dark matter}
%
In this section, we give the formulae for the elastic scattering cross 
section of the lightest neutralino off proton $\chi p$ $\to$ $\chi p$. 
The fundamental process for this scattering is the neutralino--quark 
scattering $\chi q$ $\to$ $\chi q$. 
In the non-relativistic limit, 
the relevant effective four-fermion interaction Lagrangian reads 
\cite{Drees-Nojiri-dd}
\begin{eqnarray}
{\cal L}_{\chi q} & = & 
d_q (\bar{\chi}\gamma^{\mu}\gamma_5\chi)(\bar{q}\gamma_{\mu}\gamma_5 q)
 + f_q (\bar{\chi}\chi)(\bar{q}q). 
\label{eqn:L_eff_chi_q}
\end{eqnarray}
The coupling for the spin-dependent (SD) interaction consists of 
the t-channel Z boson exchange contribution and 
the s-channel squark exchange contribution \cite{Drees-Nojiri-dd}
\begin{eqnarray}
d_q & = & 
\frac{g^2}{4m_W^2} \frac{|N_{14}|^2-|N_{13}|^2}{2}T_{3q} 
 + 
\frac{1}{4}\sum_{i=1,2} 
\frac{|C_S^{\chi q\tilde{q}_i}|^2+|C_P^{\chi q\tilde{q}_i}|^2}{m_{\tilde{q}_i}^2-(m_{\chi}+m_q)^2}, 
\label{eqn:d_q}
\end{eqnarray}
where $T_{3q}$ represents the isospin charge of the quark. 
$m_q$, $m_{\tilde{q}_i}$, $m_{\chi}$ and $m_W$ denote the masses for
the quark, the squark, the LSP and the W boson, respectively. 
The coupling for the spin-independent (SI) interaction contains 
the t-channel neutral Higgs boson exchange contribution 
and the s-channel squark exchange contribution 
\begin{eqnarray}
f_q & = & 
\sum_{r=1}^3
\frac{C_S^{\chi \chi H^0_r}C_S^{q \bar{q} H^0_r}}{m_{H^0_r}^2}
 - \frac{1}{4}\sum_{i=1,2} 
\frac{|C_S^{\chi q\tilde{q}_i}|^2-|C_P^{\chi q\tilde{q}_i}|^2}{m_{\tilde{q}_i}^2-(m_{\chi}+m_q)^2}. 
\label{eqn:f_q}
\end{eqnarray}
In general, the effective Lagrangian of neutralino--quark interaction
(\ref{eqn:L_eff_chi_q}) contains other terms: 
$(\bar{\chi}\gamma^{\mu}\gamma_5\chi)(\bar{q}\gamma_{\mu}q)$, 
$(\bar{\chi}\chi)(\bar{q}\gamma_5 q)$, 
$(\bar{\chi}\gamma_5\chi)(\bar{q}q)$ and 
$(\bar{\chi}\gamma_5\chi)(\bar{q}\gamma_5 q)$. 
However, in the non-relativistic limit, these operators are negligible 
compared to the contributions of eq.~(\ref{eqn:L_eff_chi_q}), 
since contributions of these operators are suppressed by the small 
velocity of the neutralino and the proton \cite{DD-old,dd-with-CPV}. 
We ignore them in the present analysis.

One has to convert the quark level Lagrangian (\ref{eqn:L_eff_chi_q})
to the hadronic effective Lagrangian. 
The neutralino-proton SD interaction can be found to be described by 
the following effective Lagrangian:
\begin{eqnarray}
{\cal L}_{\chi p}^{\rm SD} & = & 
d_p (\bar{\chi}\gamma^{\mu}\gamma_5\chi)(\bar{p}\gamma_{\mu}\gamma_5 p). 
\label{eqn:L_eff_chi_p}
\end{eqnarray}
The expression for the coupling constant $d_p$ is 
\begin{eqnarray}
d_p & = & \sum_{q=u,d,s} d_q \Delta_q^{(p)}. 
\label{eqn:d_p}
\end{eqnarray}
Here the parameters $\Delta_q^{(p)}$ are defined by 
$\langle p|\bar{q}\gamma_{\mu}\gamma_5q |p\rangle$ 
$=$ $2s_{\mu}\Delta_q^{(p)}$, where $s_{\mu}$ is the spin of the
proton. 
In the present analysis, we take \cite{delta_q}
\begin{eqnarray}
\Delta_u^{(p)} = 0.74, \ \ \Delta_d^{(p)} = -0.4, \ \ 
\Delta_s^{(p)} = -0.12. 
\end{eqnarray}
The SD cross section for the elastic scattering of the neutralino
can be written as 
\begin{eqnarray}
\sigma_{\chi p}^{\rm SD} & = & \frac{12}{\pi}m_r^2 d_p^2, 
\label{eqn:sigma_p_SD}
\end{eqnarray}
where $m_r$ $=$ $m_\chi m_p/(m_\chi+m_p)$ is the reduced mass.

The neutralino-proton SI interaction can be described by 
the following effective Lagrangian:
\begin{eqnarray}
{\cal L}_{\chi p}^{\rm SI} & = & f_p (\bar{\chi}\chi)(\bar{p}p). 
\label{eqn:L_eff_chi_p}
\end{eqnarray}
The coupling constant $f_p$ is given by 
\cite{SVZ,Chengs}
\begin{eqnarray}
\frac{f_p}{m_p} & = & \sum_{q=u,d,s} f_{Tq}^{(p)} \frac{f_q}{m_q}
+ \frac{2}{27}f_{TG}^{(p)} \sum_{q=c,b,t} \frac{f_q}{m_q},  
\label{eqn:f_p}
\end{eqnarray}
where $f_{TG}^{(p)}$ $=$ $1 - \sum_{q=u,d,s}f_{Tq}^{(p)}$, and 
the parameters $f_{Tq}^{(p)}$ are defined by 
$\langle p | m_q \bar{q}q |p \rangle$ $=$ 
$m_p f_{Tq}^{(p)}$ \ ($q=u,d,s$). 
In our numerical analysis, we take \cite{f_Tq}
\begin{eqnarray}
f_{Tu}^{(p)} = 0.020, \ \ f_{Td}^{(p)} = 0.026, \ \ f_{Ts}^{(p)} = 0.118. 
\end{eqnarray}
We do not include the contributions of twist-2 operators 
\cite{Drees-Nojiri-dd}, since they are typically subdominant. 
The SI cross section for the elastic 
scattering of the neutralino can be expressed as 
\begin{eqnarray}
\sigma_{\chi p}^{\rm SI} & = & \frac{4}{\pi}m_r^2 f_p^2. 
\label{eqn:sigma_p_SI}
\end{eqnarray}

%
%
\section{Relic density of the neutralino}
%
In this section, we briefly describe our method to compute 
the neutralino relic density in the MSSM with supersymmetric 
CP violation \cite{Kolb-Turner,Jungman-elal}. 

We need to evaluate the relic density at present, starting from thermal
equilibrium in the early universe. 
The time evolution of the neutralino number density $n_\chi$ 
in the expanding universe is described by the Boltzmann equation
\begin{eqnarray}
\frac{d n_\chi}{dt} + 3 H n_\chi & = & 
- \langle\sigma_{\chi\chi} v_{\rm M\o l}\rangle 
\left[ n_\chi^2 - (n_\chi^{\rm eq})^2 \right], 
\label{eqn:Boltzmann-eq}
\end{eqnarray}
where $H$ is the Hubble expansion rate, and $n_\chi^{\rm eq}$ 
is the number density which the neutralino would have in thermal 
equilibrium. 
Information of the MSSM Lagrangian is encoded 
in the cross section $\sigma_{\chi\chi}$ of the neutralino
pair annihilation into the standard model particles. 
The quantity $\langle\sigma_{\chi\chi} v_{\rm M\o l}\rangle$ 
represents the thermal average of $\sigma_{\chi\chi} v_{\rm M\o l}$,
where $v_{\rm M\o l}$ is a so-called M$\o$ller velocity which can be 
identified with the relative velocity between the two colliding 
neutralinos.

In the early universe, the neutralino is assumed to be in thermal 
equilibrium where $n_\chi$ $=$ $n_\chi^{\rm eq}$. 
As the universe expands, the neutralino annihilation process freezes
out, and after that the number of the neutralinos in a comoving volume
remains constant. 
Using an approximate solution to the Boltzmann equation 
(\ref{eqn:Boltzmann-eq}), the relic density $\rho_\chi$ $=$ $m_\chi n_\chi$
at present is given by
\begin{eqnarray}
\rho_\chi & = & \frac{1.66}{M_{\rm Pl}} 
\left(\frac{T_\chi}{T_\gamma}\right)^3 T_\gamma^3 
\frac{1}{ \int_0^{x_F}dx \langle\sigma_{\chi\chi} v_{\rm M\o l}\rangle },
\label{eqn:relic-density}
\end{eqnarray}
where $x$ $=$ $T/m_\chi$ is a temperature of the neutralino normalized 
by its mass. $T_\chi$ and $T_\gamma$ are the present temperatures 
of the neutralino and the photon, respectively. 
The suppression factor $(T_\chi/T_\gamma)^3$ $\approx$ $1/20$ follows 
from the entropy conservation in a comoving volume \cite{reheating_factor}. 
$M_{\rm Pl}$ denotes the Planck mass. 
$x_F$ is the value of $x$ at freeze-out, and 
is obtained by solving the following equation iteratively:
\begin{eqnarray}
x_F^{-1} & = & \ln \left( \frac{m_\chi}{2 \pi^3} \sqrt{\frac{45}{2g_* G_N}}
\langle\sigma_{\chi\chi} v_{\rm M\o l}\rangle_{x_F} x_F^{1/2} \right),
\label{eqn:freeze-out-temperature}
\end{eqnarray}
where $G_N$ is the Newton's constant, and 
$g_{*}$ ($\approx 81$) represents the effective
number of degrees of freedom at freeze-out.

Accurate calculation of the relic density requires a careful 
treatment of the thermal average in eq.~(\ref{eqn:Boltzmann-eq})
and the annihilation cross section in it. 
In literatures, expansion of the thermal average in powers of 
the temperature $\langle\sigma_{\chi\chi} v_{\rm M\o l}\rangle$ 
$\approx$ $a$ $+$ $bx$ is widely used. 
However, it is known that the expansion method causes large error 
when $\sigma_{\chi\chi}$ varies rapidly with the energy of the
neutralinos, 
hence, in gerenal, one has to use the exact thermal average written as 
an integration with respect to a Mandelstam variable \cite{Gondolo-Gelmini}
\begin{eqnarray}
\langle\sigma_{\chi\chi} v_{\rm M\o l}\rangle & = & 
\frac{1}{8 m_\chi^4 T K_2^2(m_\chi/T)} 
\int_{4 m_\chi^2}^\infty ds \, \sigma_{\chi\chi}(s) (s-4m_\chi^2)\sqrt{s}
K_1\left(\frac{\sqrt{s}}{T}\right),
\label{eqn:thermal-average}
\end{eqnarray}
where $K_i$ ($i$ $=$ 1,2) are the modified Bessel functions. 
The cross section $\sigma_{\chi\chi}(s)$ is a complicated function
of $s$ in general, so we have to evaluate the above integration 
numerically to calculate the thermal average.

In the annihilation cross section $\sigma_{\chi\chi}$ 
in eq.~(\ref{eqn:Boltzmann-eq}), 
there are a lot of final states: 
$\chi \chi$ $\longrightarrow$ $f \bar{f}$,
$H^0_r H^0_s$ ($r,s$ $=$ $1,2,3$), $H^+H^-$, $W^+W^-$, $ZZ$, 
$W^{\pm}H^{\mp}$, $ZH^0_r$ ($r$ $=$ $1,2,3$). 
Among these final states, the fermion pair final states $f \bar{f}$ 
often give the dominant contributions. However, depending on the parameters, 
the other final states can also play a considerable role 
\cite{rd-ww,Lopez-Nano-Yuan,NRR1}.

The expressions for the full cross section without CP violation 
at the tree level are found in Ref. \cite{NRR2}. 
We extend the analysis of Ref. \cite{NRR2} to incorporate 
the effects of CP violation. 
We have derived the full expressions for the neutralino 
annihilation cross section with CP-violating phases
at the tree level, taking into account the modified interactions
(\ref{eqn:int_h0_nn}) and (\ref{eqn:int_h0_ff}) in the presence of 
the mixing of the CP-even and CP-odd neutral Higgs fields.
In the present work,
we shall present our numerical results in the following. 
The analytic expressions for the annihilation cross section
and more comprehensive analyses of the relic density in the presence of
CP violation will be presented elsewhere.

In the present analyses, we neglect coannihilation effects
\cite{Griest-Seckel,Mizuta-Yamaguchi,ino-coan,stau-coan,stop-coan,NRR3,coan-recent}, 
though they are crucial when the LSP is higgsino-like or wino-like. 
The investigation of the effect of CP violation on the coannihilation
cross sections is left for future work.

%
%
\section{Numerical results}
%

In this section, we present our numerical results. We examine
the sensitivity of the SI and SD cross sections and the relic density 
on the phases in eq.(\ref{eqn:phase}). 
We search for the parameter sets allowed by 
the $2\sigma$ range of the WMAP constraint 
\begin{eqnarray}
0.0946 < \Omega_{\chi}h^2 < 0.1286, 
\label{eqn:WMAP-2sigma}
\end{eqnarray}
and present typical results in four cases: 
bino-like LSP case, 
mixed LSP (bino-higgsino mixing) case, 
higgsino-like LSP case and wino-like LSP case \cite{wino-like}.

\begin{figure}[t]
\hspace*{0cm}
\unitlength 1mm
\epsfxsize=15.0cm
\leavevmode\epsffile{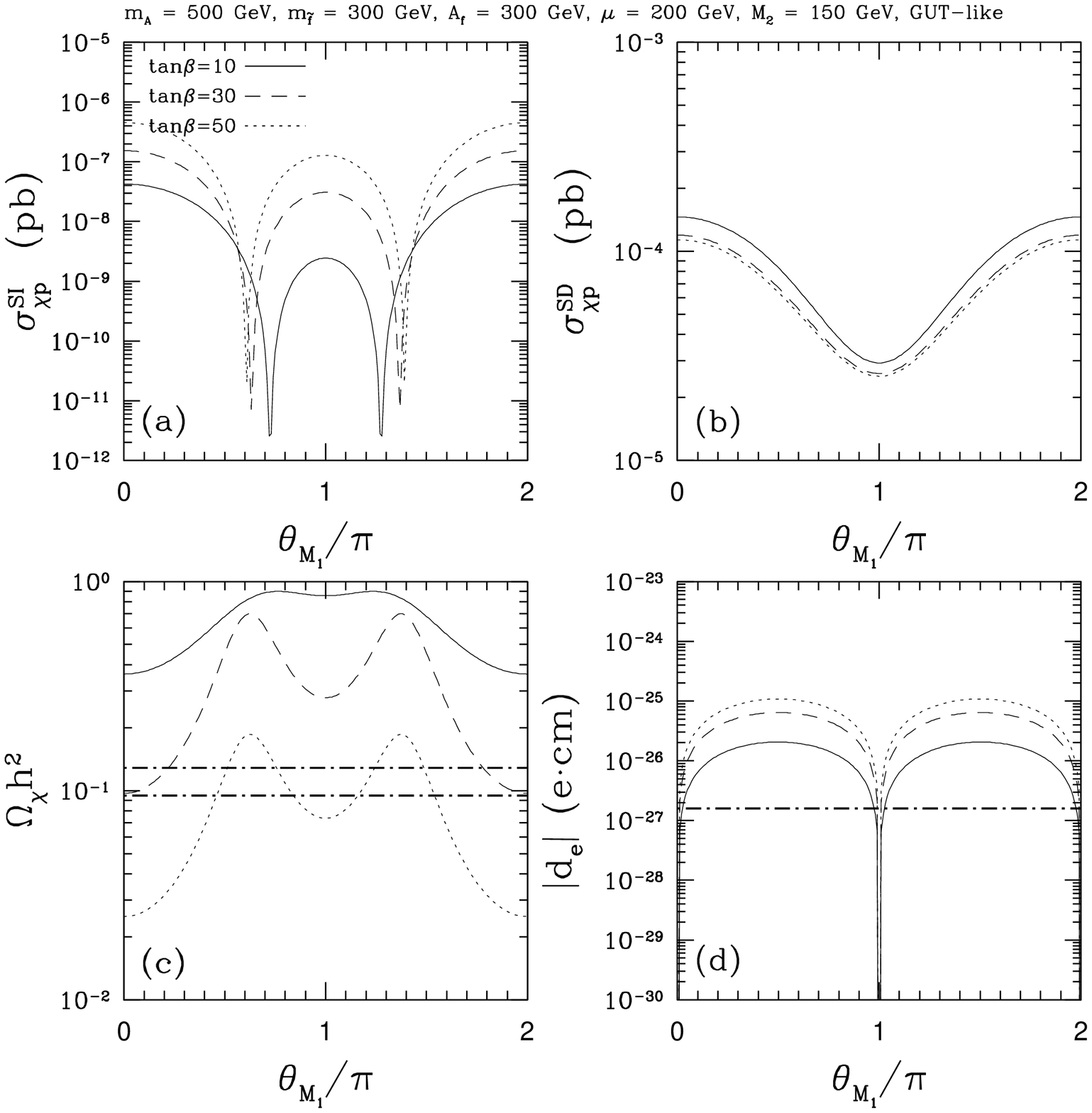}
\caption[fig1]{The results for $m_A$ $=$ $500 \gev$, $m_{\tilde{f}}$ $=$
$300 \gev$, $A_f$ $=$ $300 \gev$, $\mu$ $=$ $200 \gev$ and $M_2$ $=$
$150 \gev$ with the GUT-like gaugino mass relations (\ref{eqn:gut-like}). 
In this case, the LSP is bino-like. The spin-independent cross
section for the elastic scattering of the LSP off proton 
($\sigma_{\chi p}^{\rm SI}$), the spin-dependent one 
($\sigma_{\chi p}^{\rm SD}$), the relic density  ($\Omega_{\chi}h^2$) and 
the absolute value of the electron EDM ($|d_e|$) are plotted as
functions of the phase $\theta_{M_1}$ in windows (a), (b), (c) and (d), 
respectively.
The solid line, the dashed line and the dotted line in each window 
represent the results for $\tan\beta$ $=$ 10, 30 and 50, respectively.
In window (c), the region between the two dash-dotted lines corresponds 
to the $2\sigma$ range of the WMAP constraint $0.0946$ $<$ 
$\Omega_{\chi}h^2$ $<$ $0.1286$. In window (d), 
the region below the dash-dotted line is allowed by 
the experimental bound for the electron EDM 
$|d_e|$ $<$ $1.6 \times 10^{-27} e\cdot {\rm cm}$.}
\label{fig:bino-like-lsp-m1}
\end{figure}

Let us start with showing the effects of the phase $\theta_{M_1}$. 
One of the typical results for a bino-like LSP case is illustrated 
in Figures~\ref{fig:bino-like-lsp-m1}(a)-(d). 
In these figures, we take 
$m_{\tilde{f}}$ $=$ $300 \gev$, $A_f$ $=$ $300 \gev$,
$\mu$ $=$ $200 \gev$ and $M_2$ $=$ $150 \gev$ with
the GUT-like gaugino mass relations (\ref{eqn:gut-like}). 
The parameter $m_A$ is fixed at $m_A$ $=$ $500 \gev$ for all the 
figures in the present work. 
The solid line, the dashed line and the dotted line correspond to 
$\tan\beta$ $=$ 10, 30 and 50, respectively. 
In Fig.~\ref{fig:bino-like-lsp-m1}(a), we plot the SI neutralino-proton
cross section $\sigma_{\chi p}^{\rm SI}$ in eq.(\ref{eqn:sigma_p_SI}) 
as a function of $\theta_{M_1}$. 
There occur cancellations in the SI cross section for non-vanishing 
values of $\theta_{M_1}$ \cite{dd-with-CPV,CPJS}.

Fig.~\ref{fig:bino-like-lsp-m1}(b) shows the dependence of 
SD cross section $\sigma_{\chi p}^{\rm SD}$ in eq.(\ref{eqn:sigma_p_SD}) 
on $\theta_{M_1}$. 
The dependence is milder than that in the SI case, 
but it shows a sizable variation 
of about a factor three. The case $\theta_{M_1}$ $=\pi$ gives the
smallest value of $\sigma_{\chi p}^{\rm SD}$.

In Fig.~\ref{fig:bino-like-lsp-m1}(c), we plot the behavior of the 
neutralino relic density $\Omega_{\chi}h^2$. 
The region between the two dash-dotted lines in this figure 
corresponds to the $2\sigma$ range of the WMAP constraint 
(\ref{eqn:WMAP-2sigma}). 
For $\tan\beta$ $=$ 30 and 50, the relic density shows significant 
dependence on the phase $\theta_{M_1}$. 
In particular, we see that, for $\tan\beta$ $=$ 50, 
CP violation is necessary to find the relic density allowed by 
the WMAP constraint. 
The relic density is reduced for larger $\tan\beta$, since the 
annihilation cross section is enhanced via 
larger coupling of the neutralino with bottom-quark and bottom-squark
if the LSP includes non-zero higgsino components.

The relic density is enhanced around 
$\theta_{M_1}$ $\approx$ $2\pi/3$, $4\pi/3$ for 
$\tan\beta$ $=$ 30 and 50 in Fig.~\ref{fig:bino-like-lsp-m1}-(c). 
This enhancement results from the interferences between the major 
contributions in the neutralino annihilation cross section. 
For $\theta_{M_1}$ $=0$, the dominant contribution to 
the annihilation cross section $\sigma_{\chi\chi}$ 
comes from the bottom-quark pair creation $\chi\chi$ $\to$ $b\bar{b}$. 
In this process, a bottom-squark exchange diagram and a Higgs exchange diagram
contribute constructively for $\theta_{M_1}$ $=0$, and this leads to 
a too small relic density for $\tan\beta$ $=$ 50. 
On the other hand, for $\theta_{M_1}$ $\approx$ $2\pi/3$, $4\pi/3$ in 
the same figure, these diagrams contribute destructively. 
This interference leads to suppression of the annihilation cross
section, and the relic density is enhanced to fit into the WMAP-allowed
region for $\tan\beta$ $=$ 50.

The phase $\theta_{M_1}$ is constrained by experimentel 
constraints on the EDMs of the electron, the neutron 
\cite{EDM,Kizukuri_Oshimo,EDM_cancel} 
and 
${}^{199}{\rm Hg}$ 
\cite{mercury_EDM}. 
In the present work, we do not perform a complete analysis of the EDM 
constraints, but only present results for the electron EDM. 
In Fig.~\ref{fig:bino-like-lsp-m1}(d), we show the absolute value
of the electron EDM $d_e$. 
The current experimental bound for the electron EDM is 
$|d_e|$ $<$ $1.6 \times 10^{-27} e\cdot {\rm cm}$ \cite{d_e}. 
The upper bound is shown by the dash-dotted line in the same figure. 
With this parameter set, $|d_e|$ is beyond the upper bound for 
$\theta_{M_1}$ $\neq$ $0, \pi$. 
To satisfy the EDM constraint for $\theta_{M_1}$ $\neq$ $0, \pi$, 
we have to take heavier mass parameters. 
This typically results in a too large relic density compared to
the WMAP constraint.

\begin{figure}[t]
\hspace*{0cm}
\unitlength 1mm
\epsfxsize=15.0cm
\leavevmode\epsffile{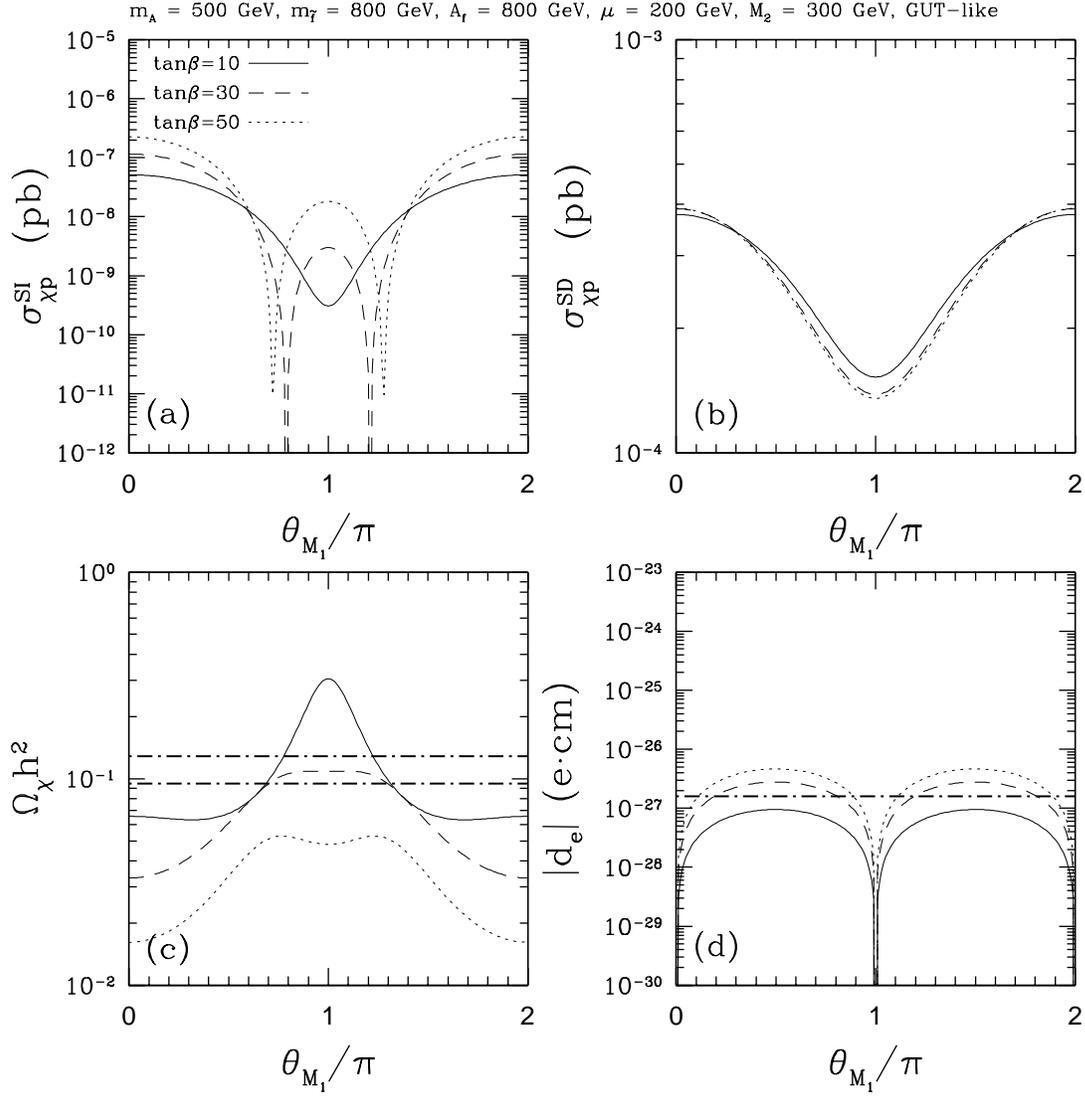}
\caption[fig1]{The same as Fig.~\ref{fig:bino-like-lsp-m1} for 
$m_A$ $=$ $500 \gev$, $m_{\tilde{f}}$ $=$ $800 \gev$, $A_f$ $=$ 
$800 \gev$, $\mu$ $=$ $200 \gev$ and $M_2$ $=$ $300 \gev$ with the 
GUT-like gaugino mass relations (\ref{eqn:gut-like}). 
This corresponds to mixed LSP with bino-higgsino mixing.}
\label{fig:mixed-lsp-m1}
\end{figure}

Figures~\ref{fig:mixed-lsp-m1}(a)-(d)
correspond to a mixed LSP (bino-higgsino mixing) case. 
In these figures, we take 
$m_A$ $=$ $500 \gev$, 
$m_{\tilde{f}}$ $=$ $800 \gev$, $A_f$ $=$ $800 \gev$,
$\mu$ $=$ $200 \gev$ and $M_2$ $=$ $300 \gev$ with
the GUT-like gaugino mass relations (\ref{eqn:gut-like}). 
The SI cross section shows cancellations at non-zero values of
$\theta_{M_1}$ for $\tan\beta$ $=$ 30 and 50. The SD cross section
is larger than the bino-like case, since for the mixed LSP case with
bino-higgsino mixing, 
either $N_{13}$ or $N_{14}$ is sizable so that 
the Z boson exchange contribution in eq.(\ref{eqn:d_q}) is increased. 
The SD cross section shows mild dependence on $\theta_{M_1}$. 
The relic density in Fig.~\ref{fig:mixed-lsp-m1}(c) again depends on 
$\theta_{M_1}$ significantly, and WMAP-allowed regions appear for
non-zero $\theta_{M_1}$ in the case of $\tan\beta$ $=$ 10 and 30. 
The WMAP-allowed regions for $\tan\beta$ $=$ 10 
also satisfy the EDM constraint as seen
in Fig.~\ref{fig:mixed-lsp-m1}(d).

\begin{figure}[t]
\hspace*{0cm}
\unitlength 1mm
\epsfxsize=15.0cm
\leavevmode\epsffile{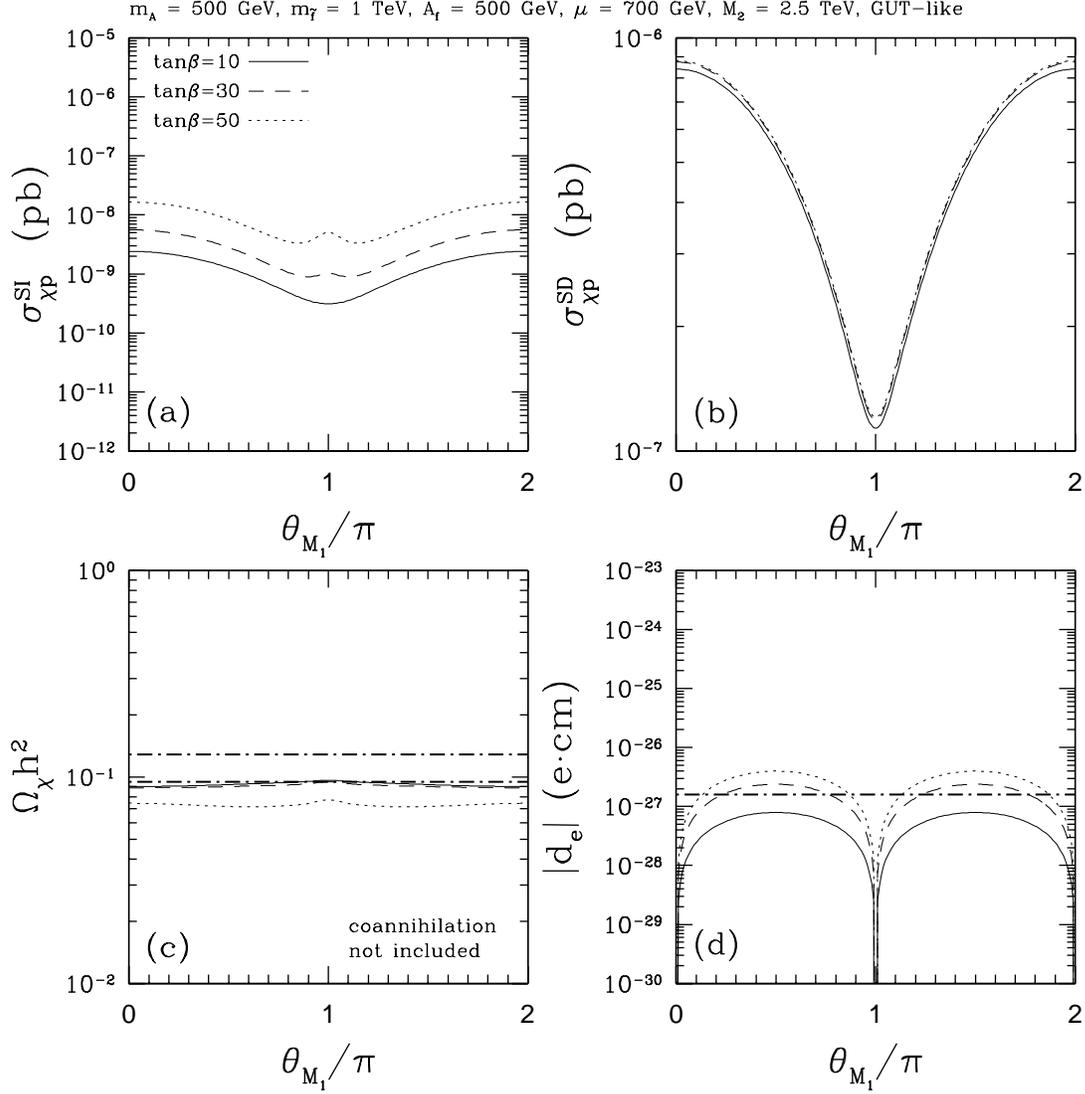}
\caption[fig1]{The same as Fig.~\ref{fig:bino-like-lsp-m1} for 
$m_A$ $=$ $500 \gev$, $m_{\tilde{f}}$ $=$ $1 \tev$, $A_f$ $=$ $500 \gev$,
$\mu$ $=$ $700 \gev$ and $M_2$ $=$ $2.5 \tev$ with
the GUT-like gaugino mass relations (\ref{eqn:gut-like}). 
The LSP in this case is higgsino-like.
Note that in our analysis on $\Omega_{\chi}h^2$, we did not include 
coannihilation effects. For the higgsino-like LSP, inclusion 
of this effects will significantly reduce $\Omega_{\chi}h^2$ so that 
the lines in window (c) will be greatly modified. }
\label{fig:higgsino-like-lsp-m1}
\end{figure}

For a higgsino-like LSP, the cross section $\sigma_{\chi\chi}$ is
relatively large, and the relic density is typically too small
to satisfy the WMAP constraint. 
In order to find WMAP-allowed regions, we have to choose relatively
large mass parameters to enhance the relic density. 
Figures~\ref{fig:higgsino-like-lsp-m1}(a)-(d)
correspond to a higgsino-like LSP case. 
In these figures, we take 
$m_A$ $=$ $500 \gev$, 
$m_{\tilde{f}}$ $=$ $1 \tev$, $A_f$ $=$ $500 \gev$,
$\mu$ $=$ $700 \gev$ and $M_2$ $=$ $2.5 \tev$ with
the GUT-like gaugino mass relations (\ref{eqn:gut-like}). 
In this case, cancellations in the SI cross section 
$\sigma_{\chi p}^{\rm SI}$ do not occur,
but $\sigma_{\chi p}^{\rm SI}$ still depends on $\theta_{M_1}$ 
as seen in Fig.~\ref{fig:higgsino-like-lsp-m1}(a). 
The SD cross section in Fig.~\ref{fig:higgsino-like-lsp-m1}(b) 
is suppressed because 
$|N_{13}|$ is nearly equal to $|N_{14}|$ for $m_W$ $\ll$ $|\mu|$ 
$<$ $|M_1|$, $M_2$ so that 
the Z boson exchange contribution in eq.(\ref{eqn:d_q}) is small. 
The relic density in Fig.~\ref{fig:higgsino-like-lsp-m1}(c) 
is not sensitive to $\theta_{M_1}$. 
In the pure higgsino limit, the mass and
couplings of the LSP 
do not include $\theta_{M_1}$ at all. This explains 
the insensitivity to $\theta_{M_1}$.

\begin{figure}[t]
\hspace*{0cm}
\unitlength 1mm
\epsfxsize=15.0cm
\leavevmode\epsffile{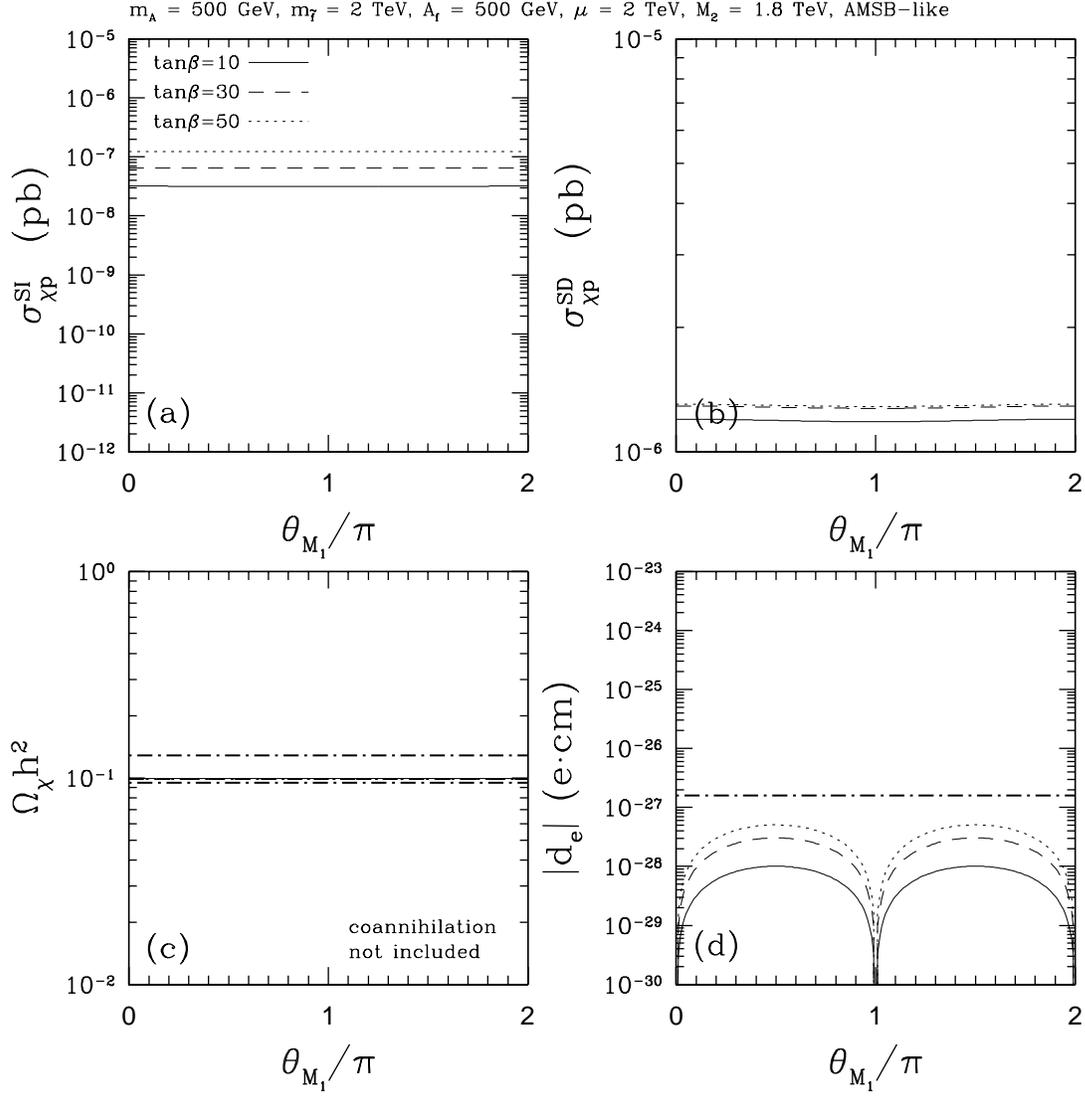}
\caption[fig1]{The same as Fig.~\ref{fig:bino-like-lsp-m1} for 
$m_A$ $=$ $500 \gev$, $m_{\tilde{f}}$ $=$ $2 \tev$, $A_f$ $=$ $500 \gev$,
$\mu$ $=$ $2 \tev$ and $M_2$ $=$ $1.8 \tev$ with
the AMSB-like gaugino mass relations (\ref{eqn:amsb-like}). 
The LSP in this case is wino-like.}
\label{fig:wino-like-lsp-m1}
\end{figure}

Figures~\ref{fig:wino-like-lsp-m1}(a)-(d)
correspond to a wino-like LSP case. 
In these figures, we take 
$m_A$ $=$ $500 \gev$, 
$m_{\tilde{f}}$ $=$ $2 \tev$, $A_f$ $=$ $500 \gev$,
$\mu$ $=$ $2 \tev$ and $M_2$ $=$ $1.8 \tev$ with
the AMSB-like gaugino mass relations (\ref{eqn:amsb-like}). 
For a wino-like LSP, the relic density is typically smaller than 
the one in a higgsino-like case so that we have to choose relatively
large mass parameters to find WMAP-allowed regions. 
We see that 
the cross sections $\sigma_{\chi p}^{\rm SI}$, $\sigma_{\chi p}^{\rm SD}$
and $\Omega_{\chi}h^2$
are not sensitive to $\theta_{M_1}$ in this case.

\begin{figure}[t]
\hspace*{0cm}
\unitlength 1mm
\epsfxsize=15.0cm
\leavevmode\epsffile{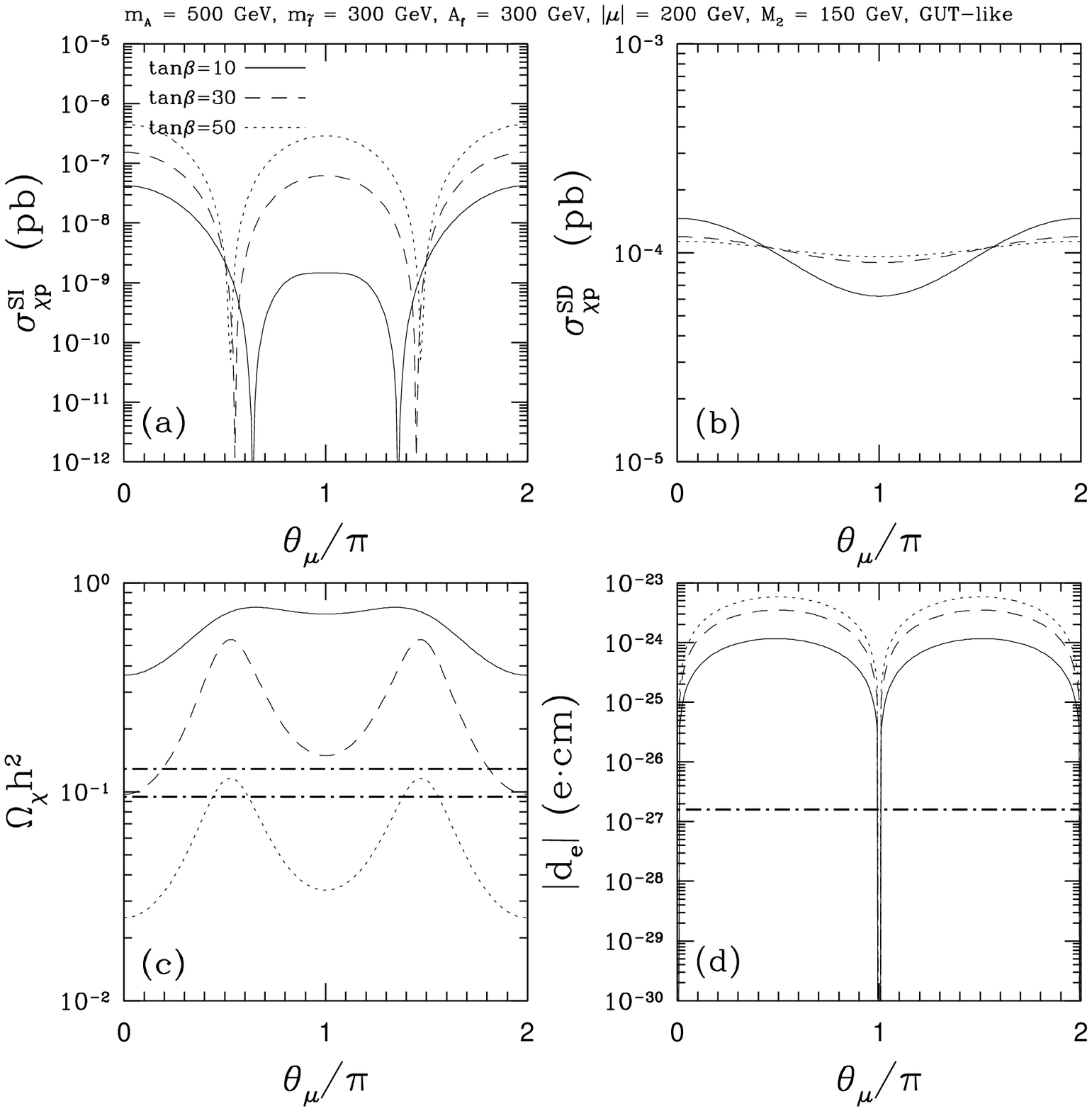}
\caption[fig1]{
The $\theta_{\mu}$ dependence of $\sigma_{\chi p}^{\rm SI}$, 
$\sigma_{\chi p}^{\rm SD}$, $\Omega_{\chi}h^2$ and $|d_e|$ 
for the same choice of parameters as in 
Fig.~\ref{fig:bino-like-lsp-m1} (the bino-like LSP case). }
\label{fig:bino-like-lsp-mu}
\end{figure}
\begin{figure}[t]
\hspace*{0cm}
\unitlength 1mm
\epsfxsize=15.0cm
\leavevmode\epsffile{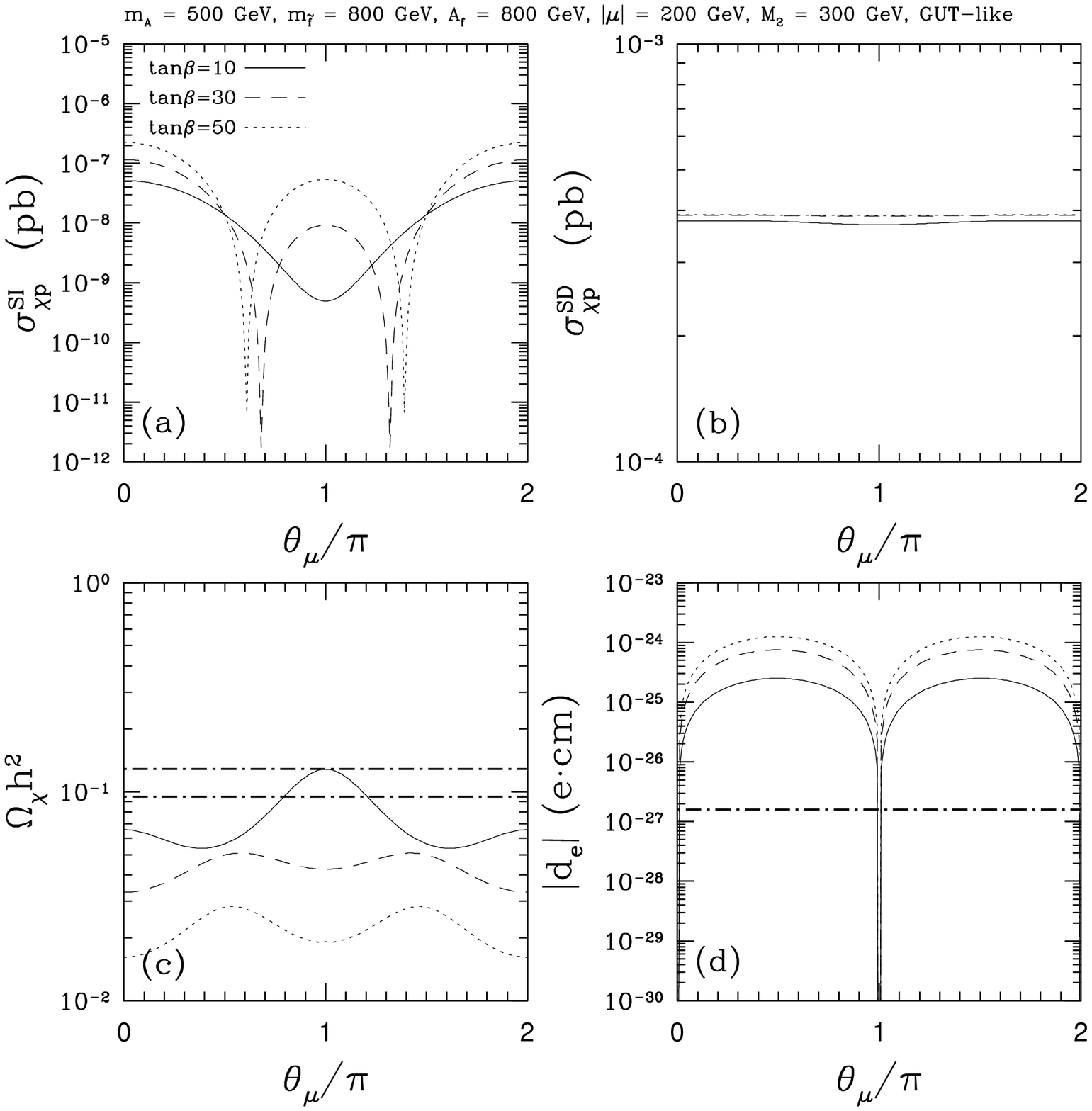}
\caption[fig1]{
The $\theta_{\mu}$ dependence of $\sigma_{\chi p}^{\rm SI}$, 
$\sigma_{\chi p}^{\rm SD}$, $\Omega_{\chi}h^2$ and $|d_e|$ 
for the same choice of parameters as in 
Fig.~\ref{fig:mixed-lsp-m1} (the mixed LSP case). }
\label{fig:mixed-lsp-mu}
\end{figure}
\begin{figure}[t]
\hspace*{0cm}
\unitlength 1mm
\epsfxsize=15.0cm
\leavevmode\epsffile{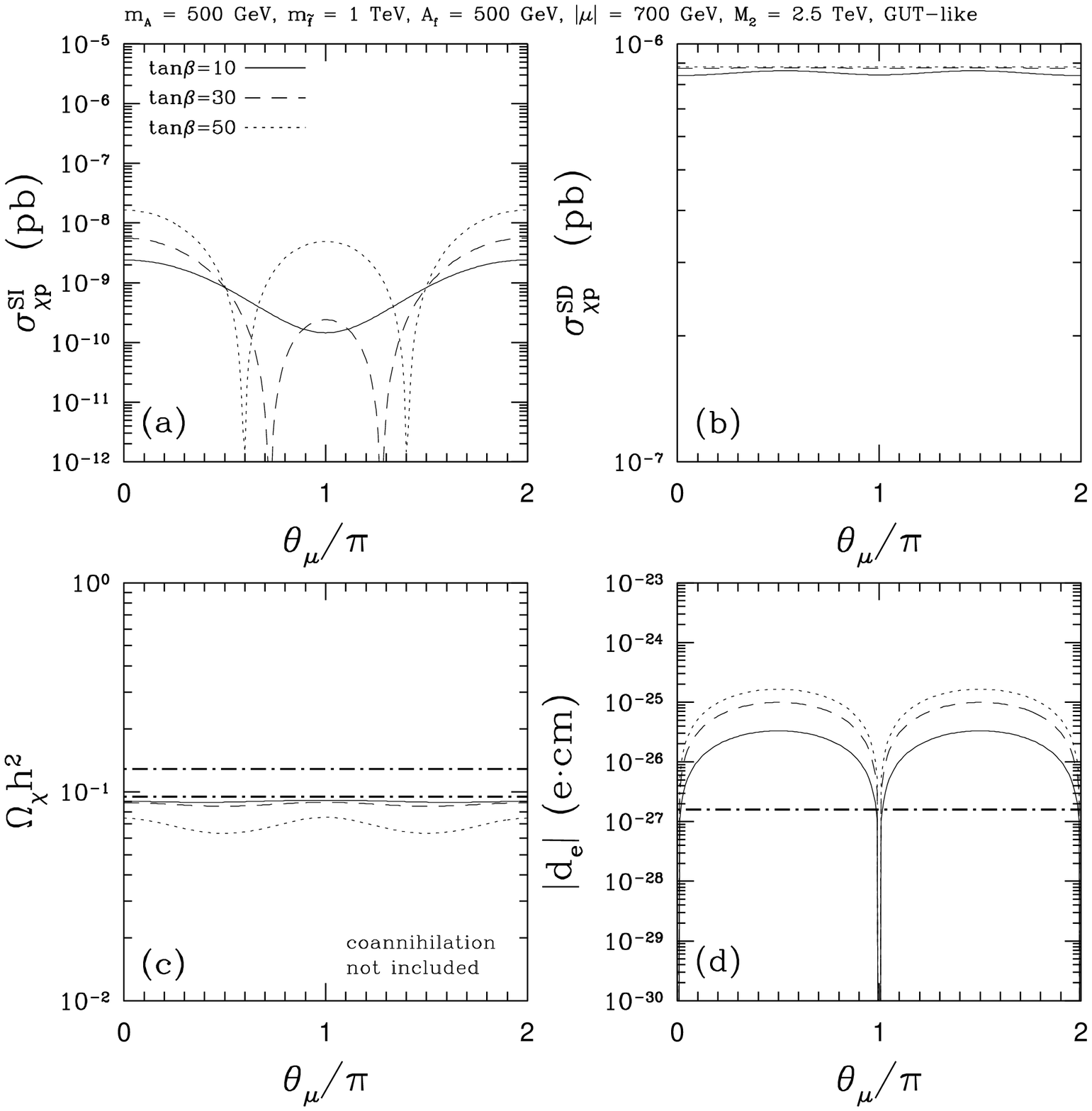}
\caption[fig1]{
The $\theta_{\mu}$ dependence of $\sigma_{\chi p}^{\rm SI}$, 
$\sigma_{\chi p}^{\rm SD}$, $\Omega_{\chi}h^2$ and $|d_e|$ 
for the same choice of parameters as in 
Fig.~\ref{fig:higgsino-like-lsp-m1} (the higgsino-like LSP case). }
\label{fig:higgsino-like-lsp-mu}
\end{figure}
\begin{figure}[t]
\hspace*{0cm}
\unitlength 1mm
\epsfxsize=15.0cm
\leavevmode\epsffile{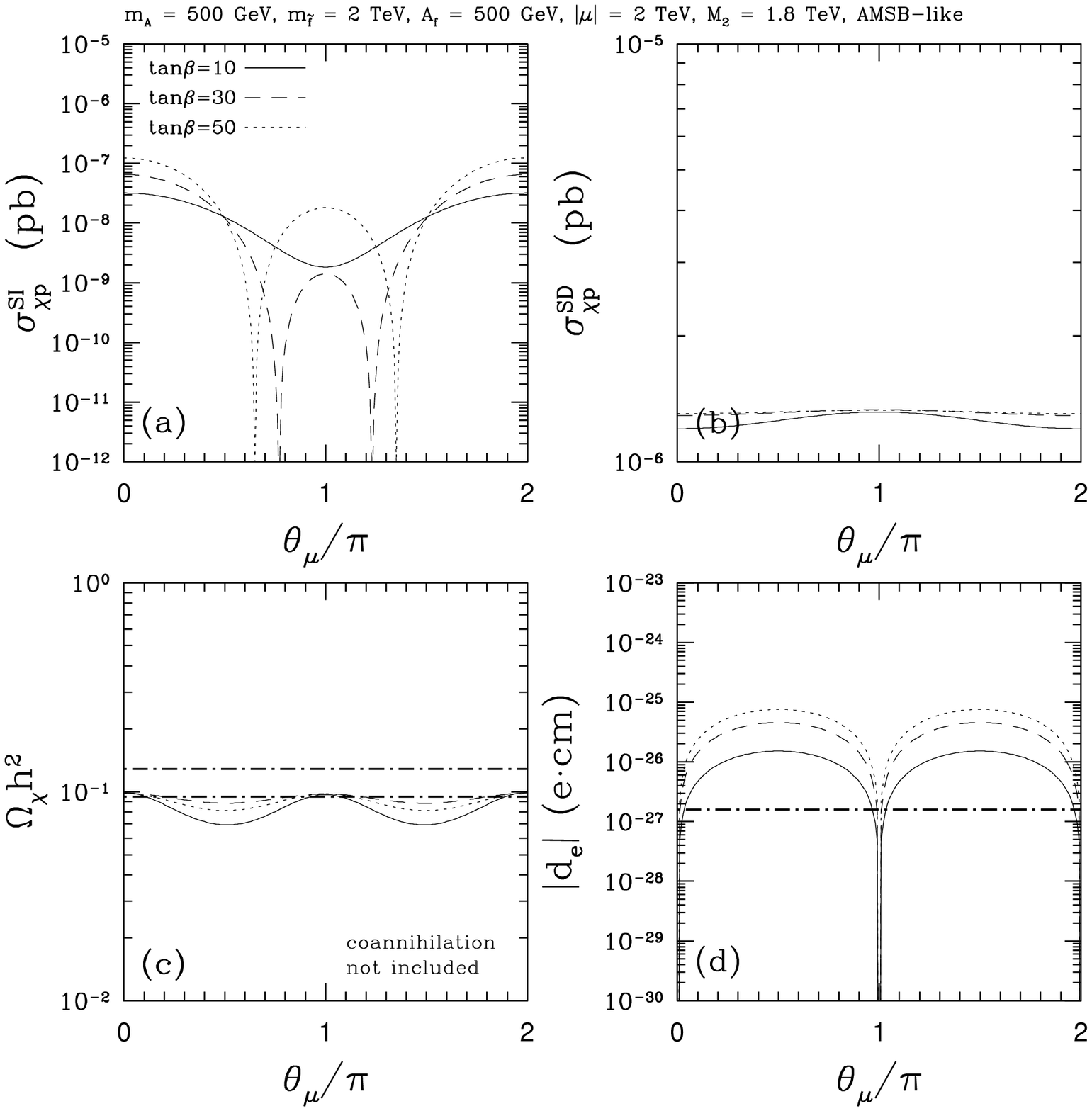}
\caption[fig1]{
The $\theta_{\mu}$ dependence of $\sigma_{\chi p}^{\rm SI}$, 
$\sigma_{\chi p}^{\rm SD}$, $\Omega_{\chi}h^2$ and $|d_e|$ 
for the same choice of parameters as in 
Fig.~\ref{fig:wino-like-lsp-m1} (the wino-like LSP case).}
\label{fig:wino-like-lsp-mu}
\end{figure}

The $\theta_{\mu}$ dependence of $\sigma_{\chi p}^{\rm SI}$, 
$\sigma_{\chi p}^{\rm SD}$, $\Omega_{\chi}h^2$ and $|d_e|$ 
is shown in Figures~\ref{fig:bino-like-lsp-mu}(a)-(d), 
\ref{fig:mixed-lsp-mu}(a)-(d), \ref{fig:higgsino-like-lsp-mu}(a)-(d) 
and \ref{fig:wino-like-lsp-mu}(a)-(d) for 
the same choice of parameters as in 
Figures~\ref{fig:bino-like-lsp-m1}(a)-(d), 
\ref{fig:mixed-lsp-m1}(a)-(d), \ref{fig:higgsino-like-lsp-m1}(a)-(d) 
and \ref{fig:wino-like-lsp-m1}(a)-(d), respectively. 
We see that the effects of $\theta_{\mu}$ are also relevant. 
Cancellations can occur in the SI cross section for non-trivial 
values of $\theta_{M_1}$. 
The SD cross section is less sensitive to $\theta_{\mu}$ compared
to $\theta_{M_1}$. 
Variations of the relic density are again significant 
for the bino-like LSP and the mixed LSP, 
while the relic density is not sensitive to $\theta_{\mu}$ for 
the higgsino-like LSP and the wino-like LSP. 
Note that the EDM constraint is much severer for $\theta_{\mu}$ than 
$\theta_{M_1}$, so that a sizable CP-violating phase $\theta_{\mu}$ 
($\neq$ 0, $\pi$) is experimentally ruled out.


In all the figures, the parameter sets do not correspond to 
resonance regions where the relic density is significantly reduced 
by resonant annihilation through the s-channel Higgs poles.
By choosing an appropriate value of $m_A$, 
we can hit the resonance regions where 
$\Omega_{\chi} h^2$ is much smaller. 
Note that, near the resonances, 
the dependence of $\Omega_{\chi} h^2$ on CP-violating 
phases through supersymmetric loop corrections to the bottom-quark 
mass is significant for large $\tan\beta$ \cite{GINS}.

In the present analysis, we have neglected coannihilation effects
\cite{Griest-Seckel,Mizuta-Yamaguchi,ino-coan,stau-coan,stop-coan,NRR3,coan-recent}. 
For Fig. \ref{fig:bino-like-lsp-m1}(c) and Fig. \ref{fig:mixed-lsp-m1}(c), 
the coannihilation effects are really negligible. 
However, for Fig. \ref{fig:higgsino-like-lsp-m1}(c) and
Fig. \ref{fig:wino-like-lsp-m1}(c), 
there exist a lighter chargino and/or a next-to-lightest neutralino
which are nearly degenerated with the LSP. 
This implies that inclusion of coannihilation effects is essential
for Fig. \ref{fig:higgsino-like-lsp-m1}(c) and
Fig. \ref{fig:wino-like-lsp-m1}(c), 
and this will result in a too small relic density compared to the 
WMAP constraint (\ref{eqn:WMAP}). 
The investigation of the effect of CP violation on the coannihilation
cross sections is left for future work.

%
%
%
%

%
%
%

%
%
\section{Conclusions}
%
We have studied 
the neutralino relic density and the neutralino-proton elastic
scattering cross sections in the MSSM with CP-violating phases. 
We have included all the final states to the neutralino pair annihilation 
cross section at the tree level, taking into account the mixing between
the CP-even and CP-odd Higgs fields. 
We have demonstrated that the variations of the relic density and the elastic 
scattering cross sections with the CP-violating phases are significant.

%
%
%
%
\section*{Acknowledgments}
The work of T.N. was supported in part by the Grant-in-Aid for Scientific
Research (No.16740150) from the Ministry of Education, Culture, Sports,
Science and Technology of Japan.'
%
%
%
%

%
%
%
%
\end{document}